\documentclass[conference]{IEEEtran}
\IEEEoverridecommandlockouts
\usepackage{bm}
\usepackage{booktabs}
\usepackage{url}
\usepackage{hyperref}
\usepackage{multirow}
\usepackage[ruled,vlined,boxed,linesnumbered]{algorithm2e}
\usepackage{threeparttable}
\usepackage{caption}
\usepackage{subcaption}
\usepackage{enumitem}

\newcommand{\tabincell}[2]{\begin{tabular}{@{}#1@{}}#2\end{tabular}}

\usepackage{cite}
\usepackage{amsmath,amssymb,amsfonts}
\usepackage{graphicx}
\usepackage{textcomp}
\usepackage{xcolor}
\def\BibTeX{{\rm B\kern-.05em{\sc i\kern-.025em b}\kern-.08em
    T\kern-.1667em\lower.7ex\hbox{E}\kern-.125emX}}
\begin{document}

\title{AutoAttention: Automatic Field Pair Selection for Attention in User Behavior Modeling}

\author{\IEEEauthorblockN{Zuowu Zheng\thanks{$^1$Z. Zheng, X. Gao, and G. Chen are with the MoE Key Lab of Artificial Intelligence,
Department of Computer Science and Engineering, Shanghai Jiao Tong University. X. Gao is the Corresponding author.}\IEEEauthorrefmark{1}, Xiaofeng Gao\IEEEauthorrefmark{1}, Junwei Pan\IEEEauthorrefmark{2}, Qi Luo\IEEEauthorrefmark{3}, Guihai Chen\IEEEauthorrefmark{1}, Dapeng Liu\IEEEauthorrefmark{2}, and Jie Jiang\IEEEauthorrefmark{2}} \IEEEauthorblockA{\IEEEauthorrefmark{1}Shanghai Jiao Tong University, Shanghai, China\\ waydrow@sjtu.edu.cn, \{gao-xf, gchen\}@cs.sjtu.edu.cn} \IEEEauthorblockA{\IEEEauthorrefmark{2}Tencent Inc., Shenzhen, China\\ \{jonaspan, rocliu, zeus\}@tencent.com}
\IEEEauthorblockA{\IEEEauthorrefmark{3}Shandong University, Shandong, China\\ luoqi2018@mail.sdu.edu.cn}}
\maketitle
\pagestyle{plain}

\begin{abstract}
In Click-through rate (CTR) prediction models, a user's interest is usually represented as a fixed-length vector based on her history behaviors. Recently, several methods are proposed to learn an attentive weight for each user behavior and conduct weighted sum pooling. 
However, these methods only manually select several fields from the target item side as the query to interact with the behaviors, neglecting the other target item fields, as well as user and context fields.
Directly including all these fields in the attention may introduce noise and deteriorate the performance. 
In this paper, we propose a novel model named \emph{AutoAttention}, which includes all item/user/context side fields as the query, and assigns a learnable weight for each field pair between behavior fields and query fields. 
Pruning on these field pairs via these learnable weights lead to automatic field pair selection, so as to identify and remove noisy field pairs. 
Though including more fields, the computation cost of AutoAttention is still low due to using a simple attention function and field pair selection.
Extensive experiments on the public dataset and Tencent's production dataset demonstrate the effectiveness of the proposed approach.

\end{abstract}


\section{Introduction}

Click-through rate (CTR) prediction is one of the most fundamental tasks for online advertising systems, and it has attracted much attention from both industrial and academic communities~\cite{chapelle2014simple,mcmahan2013ad,richardson2007predicting}. 
Modeling a user's interest through his or her history behaviors on items has proven as one of the most successful advances in the CTR prediction task~\cite{zhou2018deep,zhou2019deep,feng2019deep}.

\begin{figure}[h]
	\centering
	\includegraphics[width=\linewidth,angle=0]{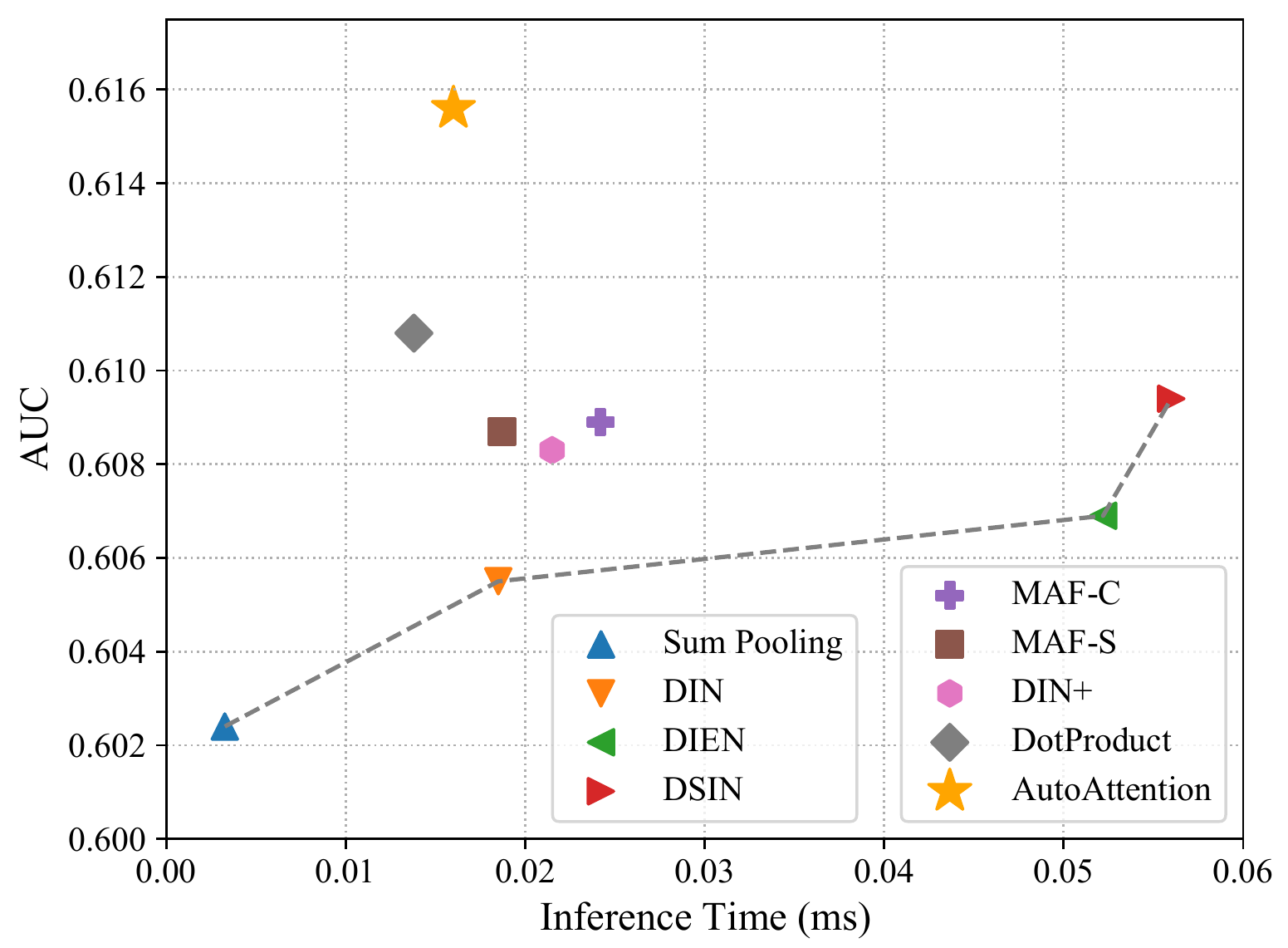}
	\caption{AUC and inference time comparison of the proposed AutoAttention with baselines on the public Alibaba dataset. Sum pooling, DIN, DIEN, and DSIN are four existing methods, which only include several manually selected fields in the attention unit. MAF-C, MAF-S, DIN+, and DotProduct are several proposed baselines which include all available fields. AutoAttention also includes all fields, but conducted field pair selection and achieves new state-of-the-art AUC with low inference time.}
	\label{fig:inference1}
\end{figure}

In the Embedding \& Multi-Layer Perceptron (MLP) algorithms for online advertising and recommendation systems, a user's interest is usually represented as a fixed-length embedding vector, based on her history behaviors~\cite{covington2016deep,zhou2018deep}.
Traditional methods take a straightforward way to do a sum or mean pooling over all behavior embedding vectors to generate one embedding~\cite{covington2016deep}. However, it ignores the fact that some behaviors are more important than others given the target item, user and context features.

Recently, several user behavior modeling methods are proposed to calculate attentive weights for different behaviors w.r.t a given target item and then conduct a weighted sum pooling, such as Deep Interest Network (DIN)~\cite{zhou2018deep} and its variants~\cite{zhou2019deep,feng2019deep}. Even though these methods achieve significant performance lift, they still suffer from the following limitations:
\begin{itemize}
    \item First, in real-world recommendation systems, a user's interest may not only depend on the target item but also on the user's demographic features or context features. However, existing works only manually select several fields from the target item side as the query and interact them with each behavior to calculate the attentive weight. It neglects the effect of other fields, including other fields from the target item side, as well as those from the user and context sides. 
    For example, when browsing the game zone of a shopping website, a boy will click a recommended new game The Witcher 3 because he clicked some similar games last week, so the item side fields should be included as all existing works do. Or it's because he is in the game zone now and any history click on games indicates a strong interest in games. In the latter case, the game zone feature from the context side plays an important role in capturing his interest from behaviors.
    \item Second, existing works interact all behavior fields with all target item side fields. Recent studies~\cite{tupe2020rethinking_position_encoding,synthesizer2021} show that some interactions in attention are unnecessary and harm the performance. Involving more fields as the query may introduce more irrelevant field interactions and further deteriorate the performance.
    \item Third, as a part of the input layer of a more complicated DNN model for CTR prediction, the procedure of generating a user interest vector should be lightweight. Unfortunately, most existing  methods use an MLP to calculate the attention weight, which leads to high computation complexity. 
\end{itemize}

To resolve these challenges, we propose to include all item/user/context fields as the query in the attention unit, and calculate a learnable weight for each field pair between user behavior fields and these query fields. To avoid introducing noisy field pairs, we further propose to automatically select the most important ones by pruning on these weights. Besides, we adopt a simple dot product function rather than an MLP as the attention function, leading to much less computation cost. We summarize the AUC as well as the average inference time of AutoAttention and several baseline models in Fig.~\ref{fig:inference1}. Except Sum Pooling which has a very low inference time due to its simplicity, the proposed AutoAttention gets a higher AUC than all the other baseline models with low inference time. The main contributions of this paper are summarized as follows:

\begin{itemize}
    \item We propose to involve all item/user/context fields as the query in the attention unit for user interest modeling. A weight is assigned for each field pair between user behavior fields and these query fields. Pruning on weights automates the field pair selection, preventing performance deterioration due to introducing irrelevant field pairs.
    \item We propose to use a simple dot product attention, rather than an MLP in existing methods. This greatly reduces the time complexity with comparable or even better performance.
    
    \item We conduct extensive experiments on public and production datasets to compare AutoAttention with state-of-the-art methods. Evaluation results verify the effectiveness of AutoAttention. We also study the learnt field pair weights and find that AutoAttention does identify several field pairs including user or context side fields, which are ignored by expert knowledge in existing works. 
\end{itemize}

The rest of the paper is organized as follows. Section~\ref{sec:pre} provides the preliminaries of existing user behavior methods. In Section~\ref{sec:model}, we describe AutoAttention, and describe its connection with several existing methods. Experiment settings and evaluation results are presented in Section~\ref{sec:exp}. Finally, Section~\ref{sec:related_works} and Section~\ref{sec:conc} discusses the related work and concludes the paper, respectively.

\section{Preliminaries}
\label{sec:pre}
\begin{figure*}[ht]
    \centering
    \includegraphics[width=\linewidth]{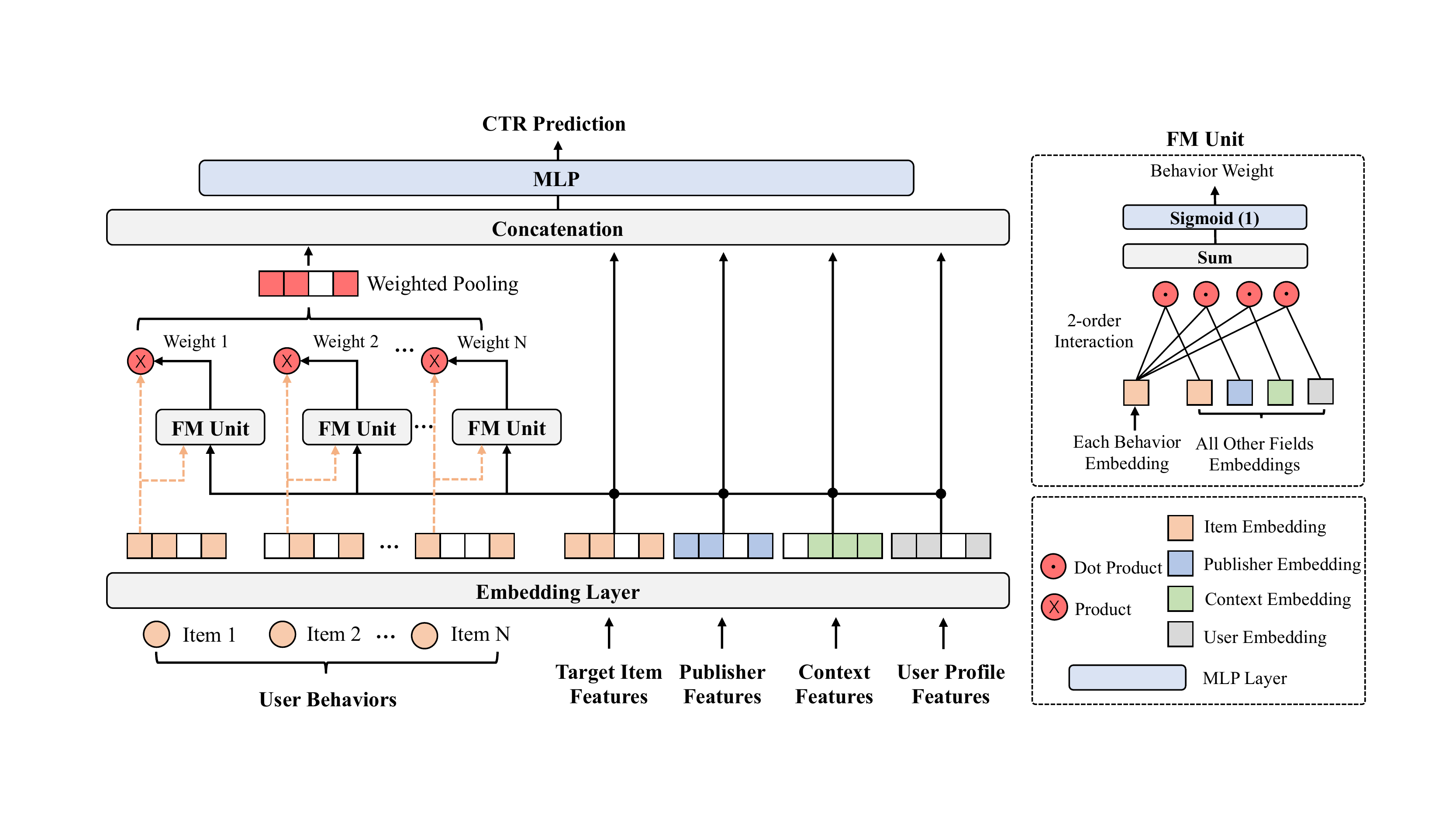}
    \caption{Architecture of the proposed AutoAttention. The input features include four parts: user behaviors, target item, user, and context. We use an attention function to calculate the weight of each behavior, the detail of which is depicted in the Attention Unit. It assigns a learnable weight for each field pair between behavior fields and query fields which consists of all item/user/context fields. Automatic field pair selection is conducted by pruning on these weights. All behaviors are summed based on the attention weights, then fed into an MLP together with all other feature embeddings.}
    \label{fig:framework}
\end{figure*}

In this section, we present the preliminaries of user behavior modeling in CTR prediction. 
A CTR prediction model aims at predicting the probability that a user clicks an item given a context (e.g., time, location, publisher information, etc.). It takes fields from three sides as the input:

\begin{equation}
    p\mbox{CTR} = f(\mbox{user}, \mbox{item},  \mbox{context})\nonumber
\end{equation}
where user side fields consists of user demographic fields and behavior fields, item and context denote fields from the item and context sides, respectively. In this paper, we focus on how to capture a user's interest from user behaviors.

Given a user $u$ and her corresponding behaviors $\{\bm{v}_1,\bm{v}_2,\dots,\bm{v}_H\}$, her interest is represented as a fixed-length vector as follows: 

\begin{equation}
    \bm{v}_u = f(\bm{v}_1,\bm{v}_2,\dots,\bm{v}_H,  \bm{e}_{F_1}, \bm{e}_{F_2}, \dots, \bm{e}_{F_M})
\end{equation}

where $\bm{v}_i$ denotes the embedding for the $i$-th behavior, $H$ denotes the length of user behaviors, and $\bm{e}_{F_j}\in \mathcal{R}^K$ denotes the feature embedding from any other field besides the user behaviors(e.g., item/user/context side fields), i.e., $F_j$. Each behavior is usually represented by multiple item side fields. Denoting the set of fields to represent behaviors as $\mathcal{B} = \{B_p\}$, then each behavior is represented as $\bm{v}_i = \sum_{B_p \in \mathcal{B}}\bm{v}_{B_p}$, where $\bm{v}_{B_p}\in \mathcal{R}^K$ denotes the feature embedding for the field $B_p$ of the $i$-th behavior.


A straightforward way to calculate $\bm{v}_u$ is to do a sum or mean pooling over all these $\bm{v}_i$ embedding vectors~\cite{covington2016deep}. However, it neglects the importance of each behavior given a specific target item.
Recently, a commonly used behavior modeling strategy is to adopt an attention mechanism over the user’s historical behaviors. It learns an attentive weight for each behavior $i$ w.r.t. a given target item $t$ and then conducts a weighted sum pooling, i.e. $\bm{v}_u = \sum_{i=1}^H a(i,t) \bm{v}_i$, where $a(i,t)$ denotes an attention function. For example, Deep Interest Network (DIN) considers the influence of the target item on user behaviors~\cite{zhou2018deep}, which learns larger weights to those behaviors that are more important given the target item, as shown in Eqn.~\eqref{dinEq}.
\begin{equation}
\begin{split}
    \bm{v}_u  &= f( \bm{v}_1, \bm{v}_2,\dots,\bm{v}_H, \bm{e}_t)\\
    &= \sum_{i=1}^H a(i,t)\bm{v}_i = \sum_{i=1}^H \text{MLP}(\bm{v}_i,\bm{e}_t)\bm{v}_i 
    \label{dinEq}
\end{split}
\end{equation}

where $\bm{e}_t$ denotes the embedding vector of the target item $t$. $\text{MLP}()$ denotes an MLP with its output as the attention weight. Following DIN, DIEN~\cite{zhou2019deep} further considers the evolution of user interest, and DSIN~\cite{feng2019deep} considers the homogeneity and heterogeneity of a user's interests within and among sessions. 
DIF-SR~\cite{dif-sr2022} proposes to only consider the interaction between corresponding fields between queries and keys within the attention.

All existing methods only interact each behavior with several selected fields from the item side within the attention, neglecting other fields, especially those from the user and context sides.


\section{AutoAttention}
\label{sec:model}

In this section, we first describe several straightforward approaches that interact user behavior with all fields in the attention unit. Then we propose AutoAttention to automatically identify and remove irrelevant field pairs which are introduced to the model due to including all fields as the query in the attention. At last, we discuss the model complexity and its connection with several existing approaches.

Mathematically, we learn a user's interest representation $\bm{v_u}$ from her historical behaviors \emph{based on all fields about the current sample}, i.e., all available fields from target item, user and context sides.
\begin{equation}
    \begin{split}
        \bm{v}_u(\bm{x}) &= f(\bm{v}_1, \bm{v}_2, \dots, \bm{v}_H, \bm{e}_{F_1}, \bm{e}_{F_2}, \dots, \bm{e}_{F_M}) \\
        &= \sum_{i=1}^Ha(\bm{v}_i,  \bm{e}_{F_{1:M}}) \bm{v}_i 
    \end{split}
    \label{eq:general1}
\end{equation}


where $\bm{v}_i$ denotes the embedding for the $i$-th behavior, which is usually a summation of several attribute embeddings for this behavior: $\bm{v}_i = \sum_p \bm{v}_{B_p}$. And $\{\bm{e}_{F_1},\dots,\bm{e}_{F_M}\}$ denotes the set of all fields from the target item, user and context sides.

\subsection{Base Models}
\label{sec:baseModel}

Before introducing AutoAttention, we first present several straightforward approaches to interact user behaviors with all fields within attention: MLP with All fields (MAF in short) and DotProduct.


\subsubsection{MLP with All Fields}

MAF simply sums or concatenates all field embedding vectors $\bm{e}_{F_{1:M}}$, then feeds it and the behavior embedding $\bm{v}_i$ to an MLP to calculate the weight. There are two ways to construct the first layer of the MLP: element-wisely sum  $\bm{e}_{F_{1:M}}$ and $\bm{v}_i$ as a $K$ dimension vector, denoted as \textbf{MAF-S} (\textbf{M}LP with \textbf{A}ll \textbf{F}ields \textbf{S}ummed); or concatenate $\bm{e}_{F_{1:M}}$ and $\bm{v}_i$ as a $(M+1)K$ dimension vector, denoted as \textbf{MAF-C} (\textbf{M}LP with \textbf{A}ll \textbf{F}ields \textbf{C}oncatenated). Mathematically,

\begin{equation}
    \begin{split}
         a_{\text{MAF-S}}(\bm{v}_i, \bm{e}_{F_{1:M}}) &=  \text{MLP}(\bm{v}_i \oplus \bm{e}_{F_1} \oplus\bm{e}_{F_2} \oplus , \cdots, \oplus \bm{e}_{F_M})\\
          a_{\text{MAF-C}}(\bm{v}_i, \bm{e}_{F_{1:M}}) &=  \text{MLP}([\bm{v}_i, \bm{e}_{F_1}, \bm{e}_{F_2}, \cdots, \bm{e}_{F_M}])
    \end{split}
\end{equation}

where $\oplus$ denotes element-wise summation, $[\cdot]$ denotes concatenation, $\mbox{MLP}(\cdot)$ denotes an Multi-Layer Perceptron, with the last layer as a single output node activated by the softmax function. 

\subsubsection{DotProduct}

Dot product is widely used in attention models~\cite{vaswani2017attention}, and it has been proved that an MLP is hard to learn a dot product~\cite{ncf_vs_mf_revisited2020}. So we propose another base model to explicitly conduct a dot product between the user behavior embedding and the sum pooling vector over all query fields. We name it as \emph{DotProduct}, formally:

\begin{equation}
    \begin{split}
        a_{\text{DotProduct}}(\bm{v}_i, \bm{e}_{F_{1:M}}) &=  \sigma \left(b + \sum_{j=1}^M \langle \bm{v}_i, \bm{e}_{F_j} \rangle\right)\\
        &= \sigma \left(b + \langle \bm{v}_i, \sum_{j=1}^M \bm{e}_{F_j} \rangle\right)
    \end{split}
\end{equation}
where $\langle \bm{v}_i, \bm{v}_j\rangle = \sum_{k=1}^K v_{i,k} \cdot v_{j,k}$ denotes the dot product function, $\sigma(\cdot)$ denotes the softmax function, and $b$ denotes the bias term.

\subsection{AutoAttention}

In above base models, all available fields are considered as the query in the attention function. However, there are still several concerns: 
1) simply involving all fields as the query ignores the fact that some fields are more important than others when being interacted with each behavior field;
2) in real-world industry systems, the number of fields is large, including all of them as the query may increase the computation cost of the model; 
3) some field pairs are irrelevant or noisy for capturing user interests, leading to performance deterioration when included in the attention.


To tackle the above mentioned challenges, we assign a weight $R_{B_p,F_j}$ to model the interaction strength for each field pair between a behavior field $B_p$ and a query field $F_j$. These field pair wise weights are learnable and trained together with all the other parameters. We name our approach as AutoAttention. Mathematically,

\begin{equation}
    a_{\text{AutoAttention}}(\bm{v}_i,  \bm{e}_{F_{1:M}}) = \sigma\left(b + \sum_{p=1}^P \sum_{j=1}^M \langle \bm{v}_{B_p}, \bm{e}_{F_j} \rangle R_{B_p,F_j}\right)
    \label{eq:AutoAttention-fs}
\end{equation}

Directly including all query fields in the attention function and interacting all of them with the behavior fields may introduce irrelevant or noisy field pairs. In order to identify and remove them, we conduct \emph{automatic field pair selection} by pruning on the field pair wise weights $R$.  There are lots of empirical studies in weight pruning area~\cite{DBLP:conf/nips/DengZLL19,DBLP:conf/iclr/FrankleC19,deng2021deeplight}. For simplicity, we adopt a standard iterative pruning algorithm used in~\cite{deng2021deeplight}. An illustration of AutoAttention is presented in Fig.~\ref{fig:framework}.


The algorithm of pruning is depicted in Alg.~\ref{alg:fields_selection}. We first train the model a few epochs to initialize the weights $R$ and then conduct pruning on $R$ to remove those with the bottom-$S\%$ lowest magnitude values. We gradually increase the sparsity rate $S\%$ such that it increases faster in the early phase when the network is stable and slower in the late phase when it becomes sensitive. Other approaches such as regularization~\cite{liu2020autofis} can also be used here.



Existing methods heavily rely on expert knowledge on selecting relevant fields to involve them in the attention. For example, in DIN~\cite{zhou2018deep}, the authors manually select three fields from the target item side: \textit{item\_id}, \textit{shop\_id} and \textit{category\_id}. Such expert knowledge is not always feasible and accurate. AutoAttention avoids such reliance on expert knowledge by automatic field pair selection.




\begin{algorithm}[t]
	\caption{Field pairs selection training procedure}
	\label{alg:fields_selection}
	\KwIn{Field pair strength weights $R$, initialize the target sparsity rate $S$, damping ratios $D$ and $U$.}
    \textbf{Warm up} Initialize whole network by training $i$ epochs\;
    \textbf{Pruning Procedure}\;
    \For{\normalfont{iteration} $j=1,2,...$}{
    Train the model for one iteration\;
    Update the current sparsity rate $S\leftarrow S\times(1-D^{j/U})$\;
    Prune the bottom-$S$\% lowest magnitude weights in $R$\;
    }
    \textbf{Online Prediction} Use the fine-tuned sparse model to make the prediction.
\end{algorithm}

\subsection{Model Training}
Follow~\cite{zhou2018deep}, after we extract a user's interest $\bm{v}_u$, we concatenate it with all the other features and feed them into an MLP:
\begin{equation}
    \hat{y} = \mbox{sigmoid}(\mbox{MLP}([\bm{v}_u, \bm{e}_{F_1}, \bm{e}_{F_2}, \cdots, \bm{e}_{F_M}]))
\end{equation}
We then minimize the following cross-entropy loss during model training:
\begin{equation}
    L(\Theta) = -\frac{1}{N}\sum_{i=1}^N(y_i\log \hat{y}_i+(1-y_i)\log (1-\hat{y}_i)) + \lambda\left\|\Theta\right\|_2
\end{equation}
where $N$ denotes the number of training samples, $\Theta$ denotes all trainable parameters, $y_i\in\{0,1\}$ denotes the label, and $\lambda\left\|\Theta\right\|_2$ denotes the $L_2$ regularization term.

\subsection{Discussion}
\begin{table*}[h]
    \centering
    \caption{A summary of model complexities. $M$ denotes the number of query fields, $P$ denotes the number of behavior fields, $K$ denotes the dimension of embedding vectors, the number of neurons of the two-layer MLP in the attention is $d$ and 1, $H$ denotes the length of user behaviors. Note that the time complexity of $H$ behaviors includes the complexity of weight calculation and weighted sum pooling over $H$ behaviors. We list the estimated FLOPs and the number of parameters in Alibaba Dataset with experiments settings of Section~\ref{sec:exp_setting}, i.e, $M=15, P=2, K=64, d=200, H=50$.}
    \renewcommand\arraystretch{1.2}
    \begin{tabular}{c|c|c|c|c|c}
    \toprule
        Model & \tabincell{c}{Time Complexity\\(One Behavior)} & \tabincell{c}{Estimated FLOPs\\(One Behavior)} & \tabincell{c}{Time Complexity\\($H$ Behaviors)} & \#Parameters of One Behavior & Estimated \#Parameters\\
        \midrule
        Sum Pooling & $O(1)$ & 0 & $O(HK)$ & 0 & 0\\
        DIN & $O(dK^2)$ & 1,642,496 & $O(K^2+dKH)$ & $dK^2+2d+1$ & 819,601\\
        DIEN & $O(K^3+dK^2)$ & 1,695,232 & $O(K^3+dKH)$ & $dK^2+12K^2+6K+2d+1$ & 869,137\\
        DSIN & $O(H^2K+dK^2)$ & 3,380,864 & $O(H^2K+K^2+dKH)$ & $2dK^2+19K^2+8K+4d+2$ & 1,717,538\\
        \midrule
        DotProduct & $O(MK)$ & 2,112 & $O(MK+HK)$ & $1$ & 1\\
        AutoAttention & $O(PMK)$ & 5,952 & $O(HPMK)$ & $PM+1$ & 31\\
    \bottomrule
    \end{tabular}
    \label{tab:complexity}
\end{table*}

\begin{figure*}[ht]
    \centering
    \includegraphics[width=\linewidth]{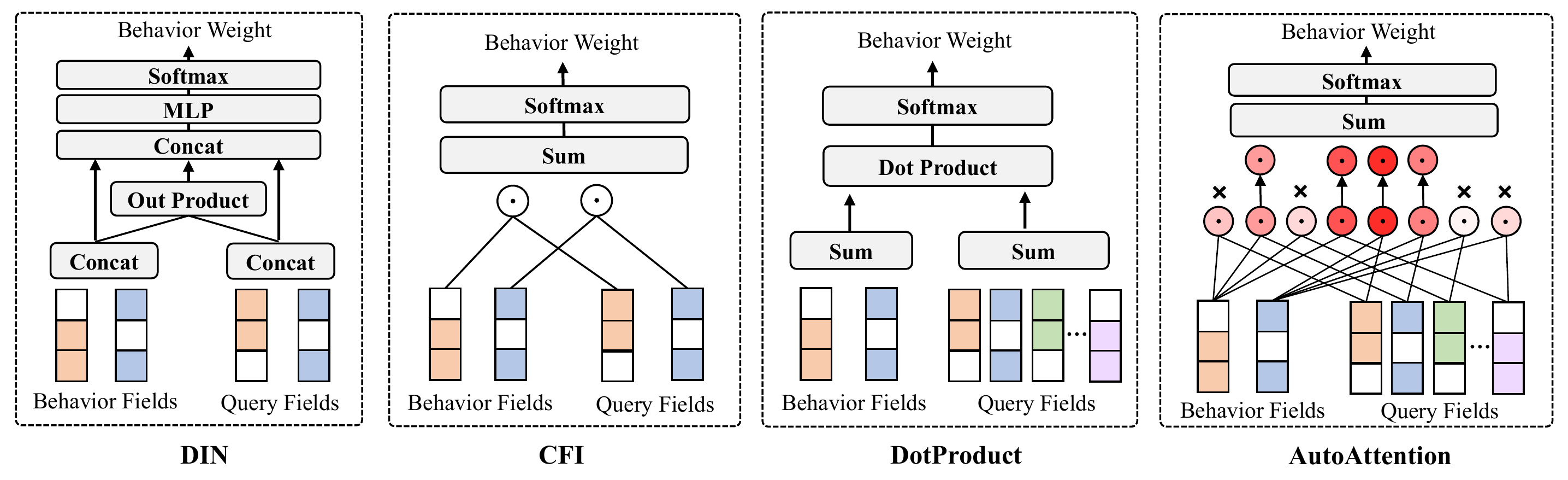}
    \caption{Model architecture comparison. DIN uses an MLP as the attention function, with several manually selected fields as the query. 
    CFI considers the corresponding field interactions between behavior fields and query fields.
    Both DotProduct and AutoAttention consider all fields in the query.
    DotProduct uses a dot product function between behaviors fields and query fields, and
    AutoAttention learns a weight $R_{B_p, F_j}$ for each behavior field and query field pair $(B_p, F_j)$. 
    The darker red color represents higher strength weights of the field pairs.}
    \label{fig:model_comparison}
\end{figure*}

\subsubsection{\textbf{Model Complexity}} The time complexity of the proposed DotProduct and AutoAttention is $O(MK)$ and $O(PMK)$ for each behavior respectively, where $M$ denotes the number of all the other fields, $P$ denotes the number of fields from the user behaviors, $K$ denotes the dimension of embedding vectors. With field pairs selection, the inference time complexity of AutoAttention can be reduced since it removes redundant field pairs during model training. Using a sparsity rate $S\%$, its inference time complexity is reduced to $O(S\%PMK)$.

As for space complexity, DotProduct only introduces one parameter, i.e., the bias term. AutoAttention introduces field pair strength weights $R$ and the bias term, which is $PM+1$. Using a sparsity rate $S\%$, AutoAttention's space complexity becomes $S\%PM+1$.
We summarize the complexities and model architectures of these models in Tab.~\ref{tab:complexity} and Fig.~\ref{fig:model_comparison}, respectively.

We also summarize the estimated number of Floating Point Operations (FLOPs), as well as the number of parameters in  Section~\ref{sec:exp_setting} from the public Alibaba dataset. As shown in Tab.~\ref{tab:complexity}, DotProduct and AutoAttention only take thousands of FLOPs to extract user interests. Compared to DNN-based methods, AutoAttention is at least two hundred of times faster and also introduces fewer parameters, which makes it a preferable choice in real-world online advertising systems.

\subsubsection{\textbf{Comparison to Self-Attention}} 

Self-attention is widely used in NLP~\cite{vaswani2017attention}, CV~\cite{DBLP:conf/iclr/DosovitskiyB0WZ21} and recommender systems~\cite{sun2019bert4rec}. In self-attention, the attention weight is a softmax over the dot product between the query $Q$ and key $K$. The value $V$ is then multiplied by these attention weights to get the final output. The proposed DotProduct can be viewed as a self-attention, taking all other fields as the query, and the user behavior as the key and value. AutoAttention further assigns a field pair wise weight for each field interaction. 
Recent works~\cite{tupe2020rethinking_position_encoding,synthesizer2021} on self-attention also reveal that some field pairs (e.g., position cross position) are critical while others are noisy within the attention, indicating the necessity of automatic field pair selection.

\subsubsection{\textbf{Comparison to CFI}}

Recently, \cite{dif-sr2022} proposes to only interact a field from behavior with the corresponding field from the target item in attention. For example, the category field of behavior is only interacted with the category field of the target item. We name this approach as CFI (Corresponding Field Interaction). CFI assumes that the corresponding field pairs are the most important ones, and all the other pairs are noisy. We compare CFI with AutoAttention in Sec~\ref{subsec:RQ2}.

\section{Experiment}
\label{sec:exp}
\begin{table*}[h]
    \centering
    \caption{Statistics of the datasets.}
    \renewcommand\arraystretch{1.1}
    \setlength{\tabcolsep}{6mm}{
    \begin{tabular}{c|c|c|c|c|c|c}
    \toprule
    Datasets & \#Train Samples & \#Test Samples & \#Fields & \#Features & \#Items & Positive Ratio\\
    \midrule
    Alibaba & 5,544,213 & 660,694 & 15 & 1,657,981 & 512,431 & 5.147\%\\
    Tencent & 6,666,928 & 1,125,130 & 15 & 1,030,047 & 388,195 & 14.167\%\\
    \bottomrule
    \end{tabular}}
    \label{tab:datasets}
\end{table*}

In this section, we evaluate AutoAttention on two real-world datasets: the public Alibaba Display Ad CTR dataset and Tencent's production CTR dataset. The code is publicly available\footnote{\url{https://github.com/waydrow/AutoAttention}}. We aim to answer the following research questions:
\begin{itemize}
    \item \textbf{RQ1}: How does AutoAttention perform compared with existing user interest methods?
    \item \textbf{RQ2}: Compared with existing methods, AutoAttention includes more fields in the attention unit, and then conducts field pair selection by pruning on field pair weights. Is it these additional fields or the field pair selection contributes more to the performance lift?
    \item \textbf{RQ3}: AutoAttention can be used as a building block of attention module in some complicated user interest methods, such as DIEN and DSIN. Does the replacement of the vanilla attention unit by AutoAttention bring performance lift?
    \item \textbf{RQ4}: Which field pairs are regarded as the most important ones in AutoAttention? Are there any important field pair regarded by AutoAttention ignored by existing methods or expert knowledge? What are these ignored field pairs?
\end{itemize}

\subsection{Datasets and Baselines}

We use the following two datasets for performance comparison. Their statistics are presented  in Tab.~\ref{tab:datasets}.

\begin{itemize}
    \item \textbf{Alibaba Dataset\footnote{\url{https://tianchi.aliyun.com/dataset/dataDetail?dataId=56}}}~\cite{feng2019deep} is a public advertising dataset released by Alibaba. It randomly samples 1,140,000 users from the website of Taobao from 8 days of click logs (26 million records) to generate the original dataset. Following~\cite{feng2019deep}, we use the first 7 days’ samples as the training set (2017-05-06 to 2017-05-12), and the next day’s samples as the testing set (2017-05-13). 
    We keep users' most recent 50 behaviors.
    Please note that we only extract the user click behaviors whose click time is before the target item to prevent information leakage.
    \item \textbf{Tencent CTR Dataset} is collected by sampling user click logs for one week from Tencent's advertising CTR log. We use samples from 2021-09-05 to 2021-09-10 as the training set, and samples on 2021-09-11 as the testing set. The data preprocessing strategy is the same as that of the Alibaba dataset.
\end{itemize}

We compare AutoAttention with the following baseline approaches:
\begin{itemize}
    \item \textbf{Sum Pooling} conducts a sum pooling without weights on the user's behavior embeddings to generate a fixed-length user interest representation.
    \item \textbf{DIN}~\cite{zhou2018deep} conducts a weighted sum pooling over user behaviors. The attention weight is calculated over the user behavior and several manually selected fields from the item side. The attention is implemented as an out product between the user behavior embedding and those selected item side embedding, followed by an MLP.
    \item \textbf{DIEN}~\cite{zhou2019deep} uses a GRU encoder to capture the behavior dependencies, followed by another GRU with an attentional update gate to depict interest evolution.
    \item \textbf{DSIN}~\cite{feng2019deep} captures users' homogeneous interests in each session and heterogeneous interests in different sessions.
    \item \textbf{GRU4Rec}~\cite{GRU4Rec} uses a GRU with ranking based loss to model user sequences for session based recommendation.
    \item \textbf{SASRec}~\cite{SASRec} uses a left-to-right Transformer to capture users' behaviors for sequential recommendation.
    \item \textbf{BERT4Rec}~\cite{Bert4Rec} uses a bi-directional self-attention to model user behaviors.
    \item \textbf{BST}~\cite{bst2019} uses self-attention and target-attention together to model user behaviors.
    \item \textbf{CFI}~\cite{tupe2020rethinking_position_encoding} considers the corresponding field interactions for sequential recommendation.
    
\end{itemize}

\subsection{Experimental Settings}
\label{sec:exp_setting}
All methods are implemented in Tensorflow 1.4 with Python 3.5, which are trained from scratch on a NVIDIA TESLA M40 GPU with 24G memory. For baseline methods, we follow the hyper-parameter settings in their original papers but also finetune them on our datasets. For both datasets, We set the maximum user behavior length $H$ to 50. The embedding dimension $K$ is 64 for all features. The dimension of each hidden layer of the three-layer MLP is 200, 80, and 1, with activation functions PReLU, PReLU, and Softmax, respectively. We use Adagrad~\cite{duchi2011adaptive} as the optimizer with a learning rate of 0.01. The batch size is 4,096 and 16,384 for the training and testing set, respectively. 
For DIN, DIEN, and DSIN, the dimension of the two-layer MLP in the local activation unit is 200 and 1, with the dice activation function~\cite{zhou2018deep}. For DSIN, we divide user behavior sequences into 5 sessions. The maximum user behavior length of each session is 10. For AutoAttention, the target sparsity rate $S$ is 0.6 and 0.8 for Alibaba and Tencent datasets respectively. The damping ratios $D$ and $U$ is set to 0.8 and 100, respectively.

We use user-weighted AUC as the evaluation metric~\cite{zhou2018deep}, which measures the goodness of samples ranking for each user. A vanilla AUC is first calculated for all samples of each user, then we conduct a weighted average over these AUCs, using the number of samples of each user as the weights. We still refer it as AUC in this paper for simplicity.
\begin{equation}
    \mbox{AUC} = \frac{\sum_{i=1}^n \#impression_i \times \mbox{AUC}_i}{\sum_{i=1}^n \#impression_i}
\end{equation}
where $n$ denotes the number of users, $\#impression_i$ and $\mbox{AUC}_i$ denote the number of impressions and AUC of the $i$-th user, respectively.

\subsection{Performance Comparison (RQ1)}

\begin{table*}[ht]
    \centering
    \caption{Experiment results of AutoAttention and baselines on the public Alibaba dataset and Tencent dataset. The bold value marks the best one in each column, while the underlined value corresponds to the second best one.}
    \renewcommand\arraystretch{1.2}
    \setlength{\tabcolsep}{3mm}{
    \begin{tabular}{c|ccc|ccc}
    \toprule
    \multirow{2}{*}{Model} & \multicolumn{3}{c|}{Alibaba} & \multicolumn{3}{c}{Tencent}\\
    & Loss (mean$\pm$std) & AUC (mean$\pm$std) & AUC Impv. & Loss (mean$\pm$std) & AUC (mean$\pm$std) & AUC Impv.\\
    \midrule
    Sum Pooling & 0.2083$\pm$0.00036 & 0.6024$\pm$0.00015 & - & 0.3561$\pm$0.00185 & 0.7125$\pm$0.00015 & -\\
    DIN & 0.2052$\pm$0.00013 & 0.6055$\pm$0.00007 & 0.515\% & 0.3545$\pm$0.00096 & 0.7173$\pm$0.00050 & 0.674\% \\
    DIEN & 0.2033$\pm$0.00020 & 0.6069$\pm$0.00062 & 0.747\% & 0.3539$\pm$0.00032 & 0.7236$\pm$0.00015 & 1.558\% \\
    DSIN & 0.2008$\pm$0.00034 & 0.6094$\pm$0.00007 & 1.162\% & 0.3526$\pm$0.00010 & 0.7285$\pm$0.00006 & 2.246\% \\
    GRU4Rec & 0.2031$\pm$0.00081 & 0.6043$\pm$0.00040 & 0.315\% & 0.3536$\pm$0.00054 & 0.7148$\pm$0.00004 & 0.323\% \\
    SAS4Rec & 0.2014$\pm$0.00043 & 0.6043$\pm$0.00015 & 0.315\% & 0.3537$\pm$0.00029 & 0.7144$\pm$0.00031 & 0.267\% \\
    BERT4Rec & 0.2027$\pm$0.00026 & 0.6049$\pm$0.00076 & 0.415\% & 0.3542$\pm$0.00016 & 0.7152$\pm$0.00063 & 0.379\% \\
    BST & 0.2016$\pm$0.00002 & 0.6050$\pm$0.00037 & 0.432\% & 0.3540$\pm$0.00005 & 0.7160$\pm$0.00052 & 0.491\% \\
    CFI & \underline{0.1983}$\pm$0.00007 & \underline{0.6115}$\pm$0.00047 & \underline{1.511\%} & \underline{0.3516}$\pm$0.00019 & \underline{0.7349}$\pm$0.00083 & \underline{3.144\%} \\
    \midrule
    AutoAttention & \textbf{0.1945$\pm$0.00062} & \textbf{0.6156$\pm$0.00053} & \textbf{2.191\%} & \textbf{0.3509$\pm$0.00081} & \textbf{0.7380$\pm$0.00040} & \textbf{3.579\%}\\
    \bottomrule
    \end{tabular}
    }
    \label{tab:exp_public}
\end{table*}


The experiment results of comparison between existing methods and our proposed AutoAttention on both datasets are shown in Tab.~\ref{tab:exp_public}. All experiments are repeated 5 times and the averaged results are reported. 

The sum pooling method is treated as a baseline. DIN gets 0.52\% and 0.67\% relative AUC lift on two datasets compared with sum pooling, since it considers the different importance of each behavior. GRU4Rec, SAS4Rec, BERT4Rec, and BST achieve similar performance, which considers sequence dependencies. DIEN and DSIN take both into account and further improve AUC. CFI gets 1.51\% and 3.14\% AUC lift respectively, which shows the advantage of corresponding field interaction.


AutoAttention significantly lifts the AUC on two datasets by 2.19\% and 3.58\%, respectively. Please note that even a 0.1\% AUC lift is huge and usually leads to a decent Gross Merchandise Volume (GMV) lift in online advertising systems~\cite{cheng2016wide}.

\subsection{Study of AutoAttention (RQ2)}
\label{subsec:RQ2}

So far, all baselines only consider several item side fields as specified in their papers, while the proposed AutoAttention uses all fields as the query and then conducts field pair selection. One may wonder whether the performance lift is mainly due to including more fields, or due to the selection.

\subsubsection{Effect of additional fields} 
To answer this question, we first equip several baseline models with all fields. Specifically, we consider all fields in three baseline models: DIN, DIEN, and DSIN, making them consist of the same fields with AutoAttention. We denote these three baselines with all fields as DIN+, DIEN+, and DSIN+. We also present the performance of the three baselines DotProduct, MAF-S, and MAF-C that already consider all fields, and AutoAttention-w/oP as a variant of AutoAttention without pruning on field pairs. The results are summarized in Tab.~\ref{tab:exp_fields}.

DIN+, DIEN+, and DSIN+ get some performance lifts compared to their original model, due to the inclusion of additional fields. For example, DSIN+ improves AUC by 0.0021 and 0.0029 on two datasets. MAF-S and MAF-C get comparable performance with DIEN+. However, they are still worse than DotProduct and AutoAttention-w/oP. AutoAttention-w/oP achieves the best result among these baselines. It indicates that explicit field interaction strength modeling is necessary between user behavior fields and query fields.

\subsubsection{Effect of \textbf{automatic} field pair selection}

CFI also conducts field pair selection by manually selecting corresponding field pairs, e.g., only interacting behavior category field with target item category field. We compare it with AutoAttention to investigate the effect of \emph{automatic} field pair selection. To make a fair comparison, we keep the same number of field pairs in AutoAttention with CFI, naming it as AutoAttention-. Please note that CFI selects $P$ fields from the target item side, and then interacts each behavior field $B_p \in \mathcal{B}$ with one of the corresponding fields among the selected ones. Therefore, AutoAttention- only keeps the top-$P$ field pairs, where $P=2$ in the Alibaba dataset and $P=3$ in the Tencent dataset.

As shown in Tab.~\ref{tab:exp_fields}, \text{AutoAttention}- performs  better than CFI. Furthermore, the relative lift of AutoAttention over CFI is 0.67\% and 0.42\%. We also compare the selected field pairs of these two methods in the following two ways: a) Compare the selected top-$P$ field pairs from AutoAttention with the $P$ corresponding field pairs in CFI. In Alibaba dataset, CFI considers two field pairs, \emph{(cate\_id, cate\_id)} and \emph{(brand, brand)}, where the first field in each pair is a query field, and the second one is a behavior field. However, \text{AutoAttention}- selects \emph{(cate\_id, brand)} and \emph{(brand, brand)}. This indicates that automatic field pair selection does not always rank the corresponding field pairs the highest; b) We check the rank of the corresponding field pairs according to the strength weight of AutoAttention. \emph{(brand, brand)} ranks 2nd and \emph{(cate\_id, cate\_id)} ranks 4th in AutoAttention. This indicates that the corresponding field pairs are indeed important according to AutoAttention, but not always the most important ones. 

\begin{table}[h]
    \centering
    \caption{Performance comparison of all baseline models using all fields and field pair selection.}
    \scalebox{1.0}{
    \renewcommand\arraystretch{1.2}
    \setlength{\tabcolsep}{0.4mm}{
    \begin{tabular}{c|cc|cc}
    \toprule
    \multirow{2}{*}{Model} & \multicolumn{2}{c|}{Alibaba} & \multicolumn{2}{c}{Tencent}\\
     & Loss & AUC & Loss & AUC\\
    \midrule
    DIN+ & 0.2020\tiny$\pm$0.00027 & 0.6083\tiny$\pm$0.00013 & 0.3541\tiny$\pm$0.00064 & 0.7196\tiny$\pm$0.00025\\
    DIEN+ & 0.2011\tiny$\pm$0.00037 & 0.6092\tiny$\pm$0.00046 & 0.3530\tiny$\pm$0.00011 & 0.7268\tiny$\pm$0.00017\\
    DSIN+ & 0.1984\tiny$\pm$0.00032 & 0.6115\tiny$\pm$0.00058 & 0.3515\tiny$\pm$0.00063 & 0.7314\tiny$\pm$0.00014\\
    MAF-S & 0.2015\tiny$\pm$0.00023 & 0.6087\tiny$\pm$0.00043 & 0.3531\tiny$\pm$0.00021 & 0.7269\tiny$\pm$0.00032\\
    MAF-C & 0.2013\tiny$\pm$0.00015 & 0.6089\tiny$\pm$0.00046 & 0.3529\tiny$\pm$0.00060 & 0.7273\tiny$\pm$0.00031\\
    DotProduct & 0.1992\tiny$\pm$0.00092 & 0.6108\tiny$\pm$0.00012 & 0.3518\tiny$\pm$0.00048 & 0.7312\tiny$\pm$0.00031\\
    \scriptsize AutoAttention-w/oP & 0.1969\tiny$\pm$0.00038 & 0.6134\tiny$\pm$0.00006 & 0.3512\tiny$\pm$0.00031 & 0.7369\tiny$\pm$0.00012\\
    \midrule
    CFI & 0.1983\tiny$\pm$0.00007 & 0.6115\tiny$\pm$0.00047 & 0.3516\tiny$\pm$0.00019 & 0.7349\tiny$\pm$0.00083\\
    \text{AutoAttention}- & 0.1972\tiny$\pm$0.00029 & 0.6124\tiny$\pm$0.00025 & 0.3515\tiny$\pm$0.00012 & 0.7366\tiny$\pm$0.00034\\
    AutoAttention & \textbf{0.1945\tiny$\pm$0.00062} & \textbf{0.6156\tiny$\pm$0.00053} & \textbf{0.3509\tiny$\pm$0.00081} & \textbf{0.7380\tiny$\pm$0.00040}\\
    \bottomrule
    \end{tabular}
    }
    }
    \label{tab:exp_fields}
\end{table}


\subsubsection{Effect of different sparsity rates}
\begin{figure}[!htbp]
	\begin{subfigure}[b]{0.5\linewidth}
		\centering
		\includegraphics[width=\linewidth]{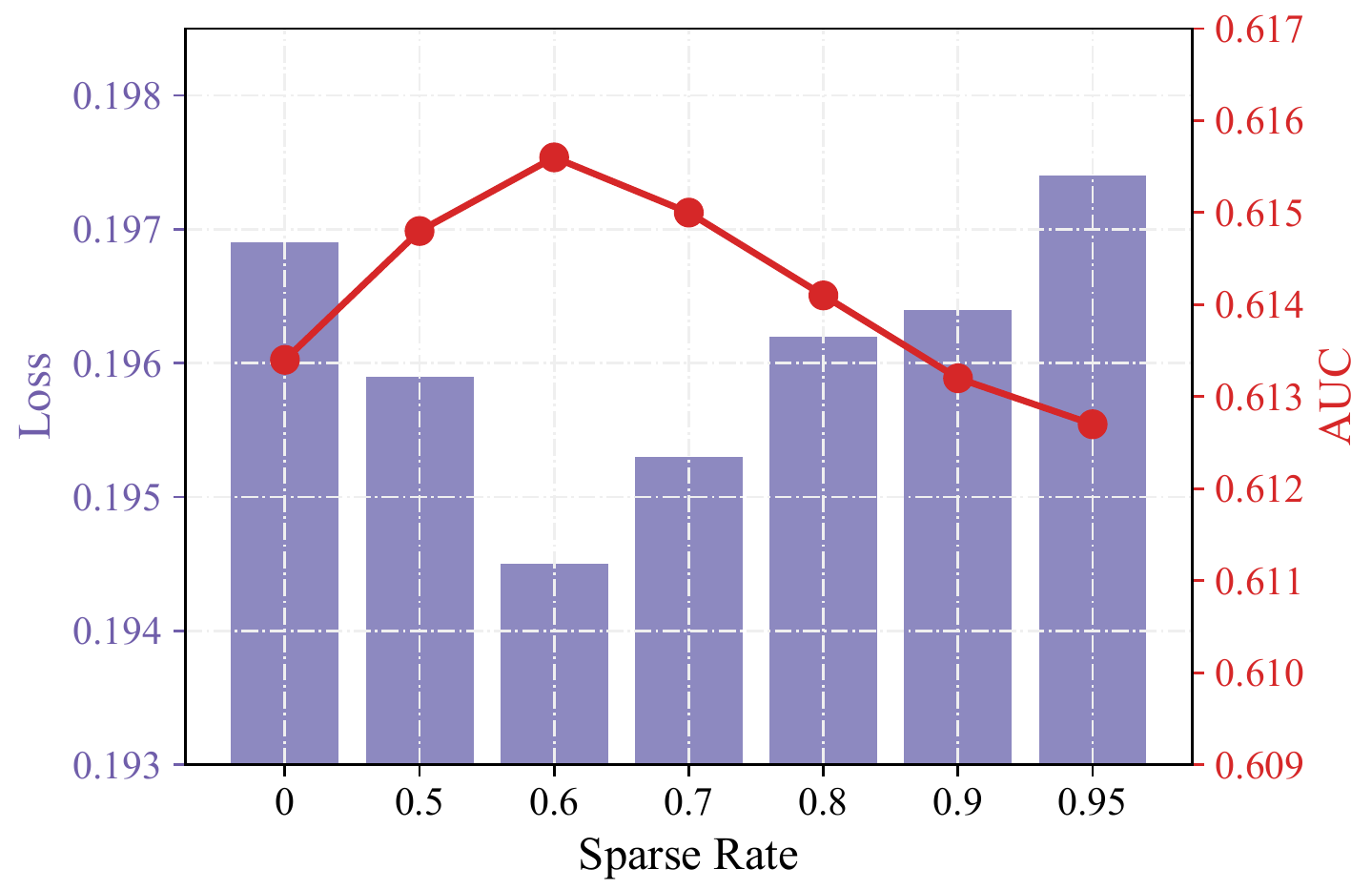}
		\caption{Alibaba Dataset}
		\label{fig:sp1}
	\end{subfigure}%
	\begin{subfigure}[b]{0.5\linewidth}
		\centering
		\includegraphics[width=\linewidth]{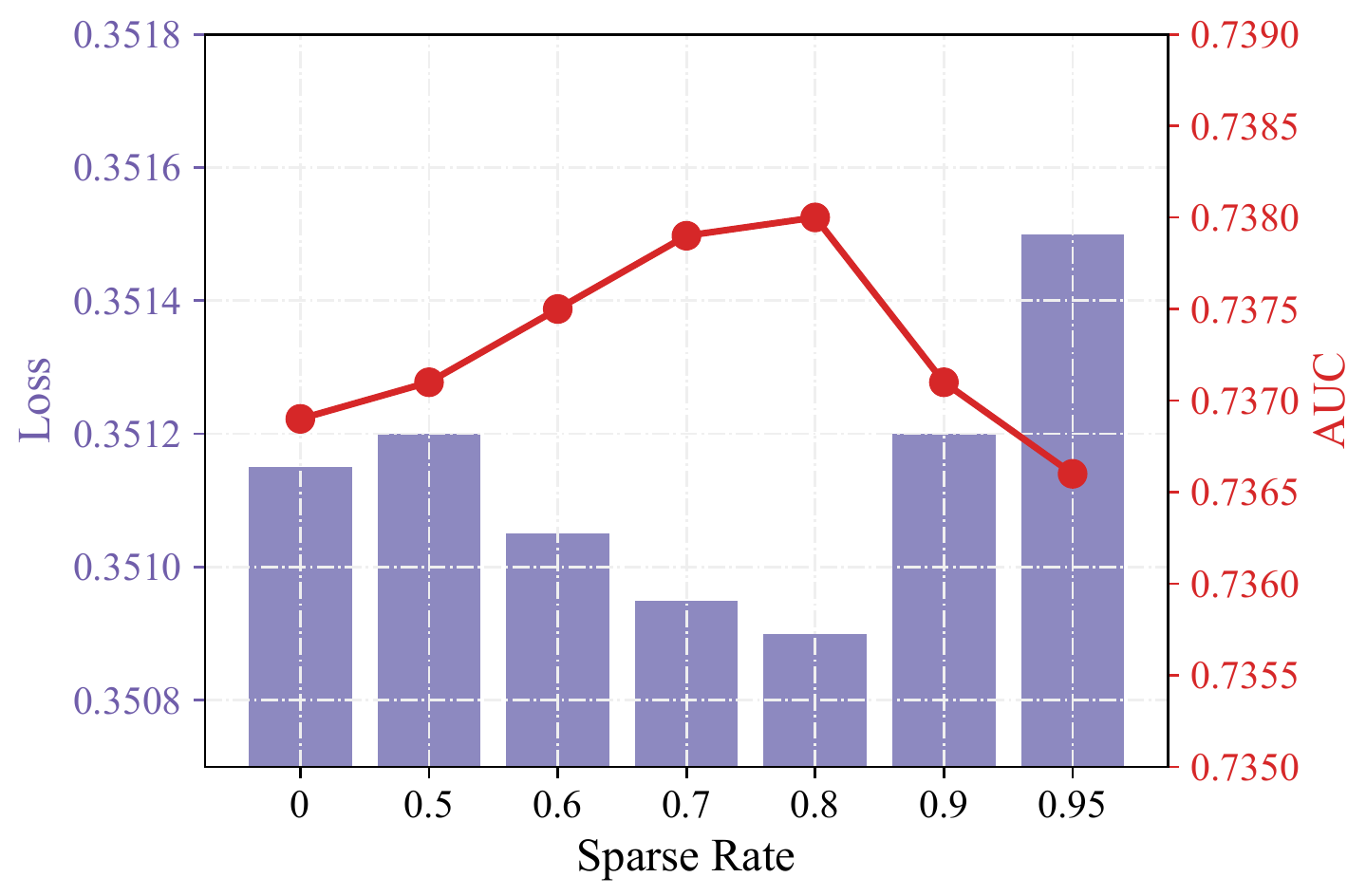}
		\caption{Tencent Dataset}
		\label{fig:sp2}
	\end{subfigure}%
	\caption{Effect of different sparsity rates $S$ on the field pair weights on two datasets.}
	\label{fig:sp}
\end{figure}

In this section, we study the impact of pruning with various sparsity rates in AutoAttention. The results are shown in Fig.~\ref{fig:sp}.

At the beginning (the leftmost), AutoAttention with $S\%=0$ means that we just assign a different weight to each field pair, but do not prune any field pair. Its performance  increases with a higher sparsity rate, and gets the best results with $S\%=0.6$ and $S\%=0.8$ on two datasets, respectively. 
The AUC lift is 0.36\% and 0.15\% compared to no pruning. This verifies that identifying and removing irrelevant field pairs leads to performance lift. With the sparsity rate becoming higher, the performance deteriorates rapidly, due to the pruning some important field pairs.




\subsection{AutoAttention as a Building Block (RQ3)}
\begin{table}[ht]
    \centering
    \caption{Experiments of replacing the original attention unit by DotProduct and AutoAttention in DIEN and DSIN.}
    \renewcommand\arraystretch{1.2}
    \setlength{\tabcolsep}{0.4mm}{
    \begin{tabular}{c|c|cc|cc}
    \toprule
    \multirow{2}{*}{Model} & \multirow{2}{*}{Attention} & \multicolumn{2}{c|}{Alibaba} & \multicolumn{2}{c}{Tencent}\\
     & & Loss & AUC & Loss & AUC\\
    \midrule
    \multirow{3}{*}{DIEN} & Original & 0.2033\tiny$\pm$0.00015 & 0.6069\tiny$\pm$0.00025 & 0.3539\tiny$\pm$0.00027 & 0.7236\tiny$\pm$0.00040\\
    & DotProduct & 0.2027\tiny$\pm$0.00016 & 0.6076\tiny$\pm$0.00005 & 0.3535\tiny$\pm$0.00009 & 0.7244\tiny$\pm$0.00018\\
    & AutoAttention & \textbf{0.2020\tiny$\pm$0.00023} & \textbf{0.6080\tiny$\pm$0.00032} & \textbf{0.3531\tiny$\pm$0.00010} & \textbf{0.7258\tiny$\pm$0.00035}\\
    \midrule
        \multirow{3}{*}{DSIN} & Original & 0.2008\tiny$\pm$0.00008 & 0.6094\tiny$\pm$0.00005 & 0.3526\tiny$\pm$0.00037 & 0.7285\tiny$\pm$0.00076\\
    & DotProduct & 0.2001\tiny$\pm$0.00034 & 0.6098\tiny$\pm$0.00042 & 0.3523\tiny$\pm$0.00019 & 0.7291\tiny$\pm$0.00013\\
    & AutoAttention & \textbf{0.1998\tiny$\pm$0.00070} & \textbf{0.6101\tiny$\pm$0.00047} & \textbf{0.3521\tiny$\pm$0.00068} & \textbf{0.7294\tiny$\pm$0.00062}\\
    \bottomrule
    \end{tabular}
    }
    \label{tab:exp_adaptation}
\end{table}

AutoAttention can be treated as a general purpose attention unit. In addition to using it standalone to model a user's interest, we can also replace the original attention unit within DIEN and DSIN by the proposed DotProduct and AutoAttention. Note that we can not obtain the original field embeddings of user behaviors since DIEN and DSIN use hidden states to represent behavior within attention. Therefore, we only assign field-wise weights for fields from the target item rather than the field pair wise weights.
In DIEN, we replace its attention function by AutoAttention:
\begin{equation}
    \alpha_t = \sigma\left(b + \sum_{j=1}^{M^\prime} \langle \bm{h}_t, \bm{e}_{F_j} \rangle R_{F_j}\right)
    \label{eqn:dien_2}
\end{equation}
where $M^\prime$ denotes the number of fields from the target item side, $\bm{h}_t$ is the hidden state of user behavior. While in DSIN, we replace the attention function in the session interest activating layer. The formulation is similar to Eqn.~\eqref{eqn:dien_2}, except that we use an embedding vector of session interest instead of a hidden state of user behavior.

As shown in Tab~\ref{tab:exp_adaptation}, the performance of DIEN and DSIN can be further boosted with the two proposed attention units. Note that replacing the original attention unit by AutoAttention introduces marginal additional parameters and computational cost, which can be ignored compared with the cost of the two methods themselves.


\subsection{Visualization of field pair selection (RQ4)}
\begin{figure}[!htbp]
    \centering
	\begin{subfigure}[b]{\linewidth}
		\centering
		\includegraphics[width=\linewidth]{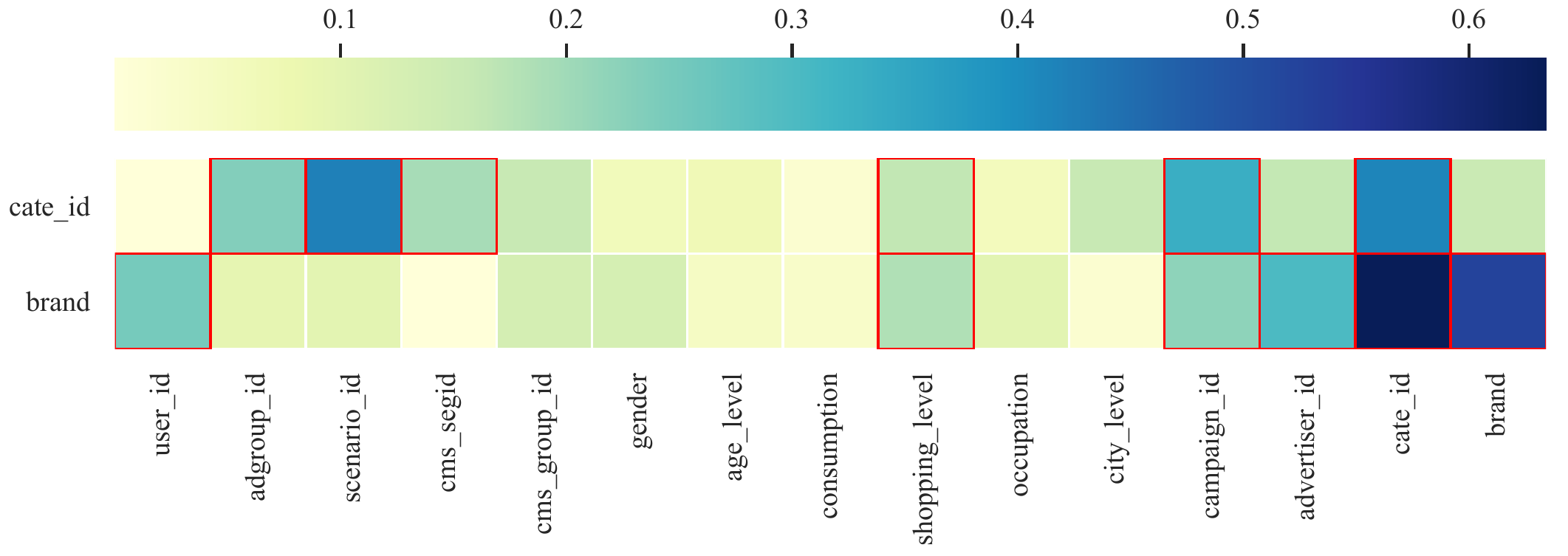}
		\caption{Alibaba Dataset}
		\label{fig:fs1}
	\end{subfigure}\\
	\begin{subfigure}[b]{\linewidth}
		\centering
		\includegraphics[width=\linewidth]{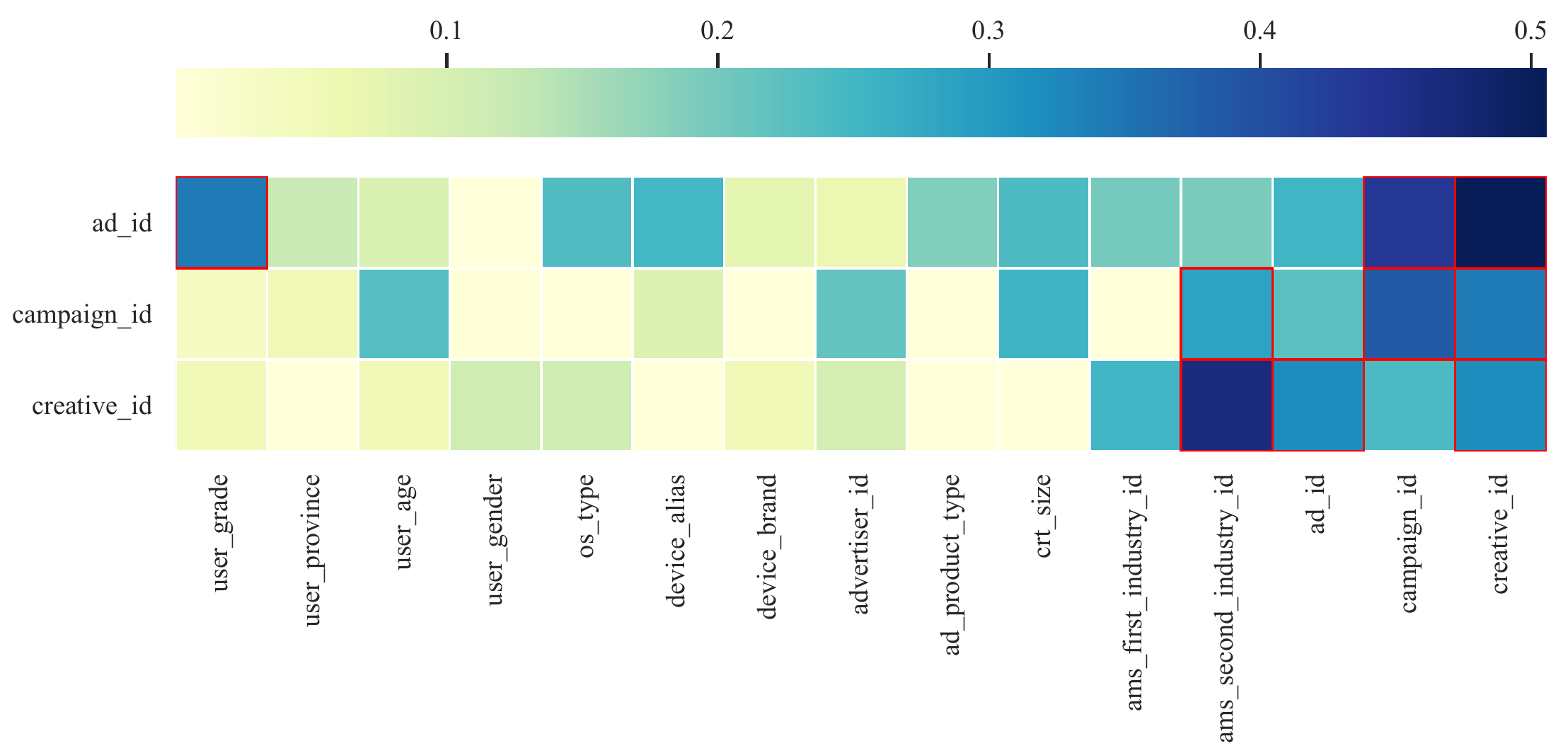}
		\caption{Tencent Dataset}
		\label{fig:fs2}
	\end{subfigure}%
	\caption{Heat map of learnt field pairs strength weights $R$ of AutoAttention on two datasets. The cells with red box denote the selected field pairs in AutoAttention.}
	\label{fig:fs}
\end{figure}

We would verify whether the learnt field pairs weights really reflect the importance of each field interaction, and whether we recognize some other important field pairs which are neglected by the expert knowledge. We visualize the learnt field pair strength weights $R$ by heat maps on two datasets, shown in Fig.~\ref{fig:fs1} and Fig.~\ref{fig:fs2}, where the  x-axis and y-axis denote the fields from current sample (including item/user/context fields) and user behaviors, respectively.

We observe that the field pairs within item side fields indeed play an important role within the attention, such as \textit{(cate\_id, brand)}, \textit{(brand, brand)}, \textit{(creative\_id, ad\_id)}, and \textit{(ams\_second\_industry\_id, creative\_id)}. In addition, we also observe that some field pairs from other sides are also important, such as \textit{(scenario\_id, cate\_id)} and \textit{(user\_grade, ad\_id)}. These field pairs are usually neglected when manually selecting fields or field pairs, which can be identified in AutoAttention.

\section{Related Works}
\label{sec:related_works}
In this section, we discuss two research areas related to our work, i.e., CTR prediction and user behavior modeling.

\subsection{CTR Prediction}
CTR Prediction is one of the most fundamental tasks in online advertising and recommendation systems, which aims at predicting the probability that a user clicks an item or ad. 
Pioneer works of CTR prediction are proposed mainly based on Logistic Regression (LR)~\cite{richardson2007predicting,chapelle2014simple,mcmahan2013ad}, polynomial~\cite{chang2010training}, collaborative filtering~\cite{shen2012personalized}, tree models~\cite{he2014practical}, Bayesian models~\cite{graepel2010web}, etc. 
In order to explicitly model the feature interactions, many factorization machine based methods are proposed for high-dimensional data, such as Factorization Machine (FM)~\cite{rendle2010factorization}, Field-aware Factorization Machine (FFM)~\cite{juan2016field}, Field-weighted Factorization Machine (FwFM)~\cite{pan2018field,pan2019predicting}, and Field-matrixed Factorization Machine (FmFM)~\cite{sun2021fm}. Besides, there are some works that aim at learning weight for different feature interactions, including Attentional Factorization Machines (AFM)~\cite{xiao2017attentional}, Dual-attentional Factorization Machines (DFM)~\cite{liu2020dual}, Dual Inputaware Factorization Machines (DIFM)~\cite{lu2020dual}.

Since the number of samples and the dimension of features have become larger and larger, many deep learning based models have been proposed, such as Wide\&Deep~\cite{cheng2016wide}, Deep Crossing~\cite{shan2016deep}, YouTube Recommendation~\cite{covington2016deep}, PNN~\cite{qu2016product}, Deep\&Cross~\cite{wang2017deep}. There are some studies that combine FM with DNN, such as DeepFM~\cite{guo2017deepfm},  NFM~\cite{he2017neural}, xDeepFM~\cite{lian2018xdeepfm}, InterAtt~\cite{li2020interpretable}, DeepLight~\cite{deng2021deeplight} and DCN V2~\cite{dcnv22021}. 


\subsection{User Behavior Modeling}
Traditional methods take a straightforward way to represent each behavior with an embedding vector, and then do a sum or mean pooling over all these embedding vectors to generate one embedding~\cite{covington2016deep}. Then many works propose to assign a dynamic weight for each behavior and then conduct weighted sum pooling, such as Deep Interest Network (DIN)~\cite{zhou2018deep}, Deep Interest Evolution Network (DIEN)~\cite{zhou2019deep}, and Deep Session Interest Network (DSIN)~\cite{feng2019deep}. There are also many works to use RNN or Transformer for behavior sequence modeling, including GRU4Rec~\cite{GRU4Rec}, SAS4Rec~\cite{SASRec}, BERT4Rec~\cite{Bert4Rec}, and Behavior Sequence Transformer (BST)~\cite{bst2019}.


Recently, there are also some works further consider long-term historical behavior sequences, such as Multi-channel user Interest Memory Network (MIMN)~\cite{pi2019practice}, Hierarchical Periodic Memory Network (HPMN)~\cite{ren2019lifelong}, Search-based Interest Model (SIM)~\cite{pi2020search}, UBR4CTR~\cite{ubr4ctr}, ETA~\cite{eta} and LimaRec~\cite{limarec}.

\section{Conclusion}
\label{sec:conc}

In this paper, we propose an efficient user interest model AutoAttention for CTR prediction. We propose to include all fields from item/user/context sides as the query fields and interact them with behavior fields within attention. We assign a learnable weight for each field pair between behavior fields and query fields to capture their different importance. Pruning fields pairs via these weights can identify and remove irrelevant and noisy field pairs, leading to performance lift and computation complexity reduction. Comprehensive experiments on public and production datasets demonstrate the effectiveness of the proposed approach.


\section*{Acknowledgment}
This work was supported by the National Key R\&D Program of China [2020YFB1707903]; the National Natural Science Foundation of China [61972254, 62272302]; Shanghai Municipal Science and Technology Major Project [2021SHZDZX0102]; the CCF-Tencent Open Fund [RAGR20200105]; and the Tencent Marketing Solution Rhino-Bird Focused Research Program [FR202001].

\bibliographystyle{IEEEtran}
\bibliography{ref}

\end{document}